\documentclass[aps,prl,twocolumn,groupedaddress,showpacs]{revtex4}
\usepackage{graphicx}

\begin{document}
\title{RamaDA: complete and automated conformational overview of proteins}
\author{Matthieu Tanty$^{1,2}$ and Marc-Andr\'e Delsuc$^1$}
\address{1) Institut de G\'en\'etique et de Biologie Mol\'eculaire et Cellulaire (IGBMC), UMR 7104, 1 rue Laurent Fries, BP 10142, 67404 Illkirch cedex, France\\
2) NMRtec, Bioparc B, boulevard S\'ebastien Brandt, 67400 Illkirch, France }
\date{\today}

\begin{abstract}

The tertiary structure of protein, as well as the local secondary structure organization are fully determined by the angles of the peptidic bound.
The backbone dihedral angles not only determine the global fold of the protein, but also the details of the local chain organization.
Although a wealth of structural information is available in different databases and numerous structural biology softwares have been developed, rapid conformational characterization remains challenging.

We present here RamaDA, a program able to give a synthetic description of the conformation of a protein.
The RamaDA program is based on a model where the Ramachadran plot is decomposed into seven conformational domains.
Within the framework of this model, each amino-acid of a given protein is assigned to one of these domains.

From this assignment secondary structure elements can be detected with an accuracy equivalent to that of the DSSP program for helices and extended strands, and with the added capability of detecting PolyProline II secondary structures.
Additionally, the determination of a z-score for each amino-acid of the protein emphasizes any irregularities in the element.

It is also possible to use this analysis to detect characteristic conformational patterns.
In the case of EF-hands, calcium-binding helix-loop-helix domains, it is possible to design a strict consensus for the 9 amino-acids of the loop.
523 calcium binding protein files can be found into the entire PDB with this pattern and only 2.7\% false positive hits are detected.

The program RamaDA gathers several tools in one and is then able to give a complete information on a protein structure, including loops and random coil regions.
Through the example of EF-hands, a promising approach of structural biology is developed.
RamaDA is freely available for download as well as online usage at \url{http://ramada.u-strasbg.fr}
\end{abstract}

\keywords{Ramachandran plane, secondary structure, EF-hand, conformational patterns }

\maketitle

%
\section*{Background}

A protein is a hierarchical molecule, with a structure tha is organized in primary, secondary and tertiary structures.
However, the tertiary structure, as well as the local secondary structure organization, are fully determined by the angles $(\varphi,\psi )$ of the peptidic bound.
The backbone dihedral angles not only dictate the global fold of the protein, but also the details of the local chain organization.

The Ramachandran plot, first proposed in 1965 by Ramakrishnan and Ramachandran \cite{Ramakrishnan:1965p211} is a tailor-made tool to study the conformations adopted by amino-acids. 
This plot uses the dihedral angles $\varphi$ et $\psi$ to indicate if a specific pair is sterically allowed and/or which conformational domain is adopted \cite{Ramakrishnan:1965p211,Edsall:1966p219}.
\emph{Allowed} (or \emph{favoured}) regions of this space have been associated with regular secondary structure elements such as $\alpha$-helix or $\beta$-sheet, while empty \emph{disallowed} regions have been highlighted.

Since Ramakrishnan and Ramachandran’s initial work, several conformational domains have been identified in the \emph{allowed} regions of the plot \cite{Kleywegt:1996p34, Hovmoller:2002p222, Lovell:2003p32}. 
In the literature, one can find the extended region that can be split into the $\beta$-sheet and the PolyProline-II (PPII) domains \cite{Munoz:1994p220},
  the $\alpha$ domain corresponding to right-handed helical conformations,
  the $\gamma$ domain corresponding to a specific conformation of hydrogen bonded $\gamma$-turns \cite{MilnerWhite:1990p230},
  the $\zeta$ domain, which is exclusively composed of conformations of amino-acids preceding a proline \cite{Ho:2005p33}, 
  the $\alpha_L$ domain corresponding to left-handed helical conformations 
  and the PPII$_R$ domain, sometimes noted $\beta_{PR}$ \cite{Ho:2005p33}, corresponding to right-handed PPII helical conformations.
The existence of these conformational domains is only based on sterical hindrance and do not take into account any other parameter or external force. 

The amount of structural information available in databases such as the Protein DataBank (PDB) \cite{Berman:2000p1167} has increased much faster than the number of programs analyzing it. 
Actually, few programs and databases can give accurate local information on proteins in the PDB \cite{Cock:2009p227,Bornot:2011fk,Joosten:2011p2532,Zimmermann:2011p2533} and it remains a challenge to get this information. 
However, given the wealth of structural information available in the PDB, it is possible to develop a statistical model of the Ramachandran plot.
From this model, we have developed a program called RamaDA (for Ramachandran Domain Analysis).

The RamaDA program takes into account all the coordinates found in the analyzed file, including the different models of the same protein created with NMR constraints, in order to assign a conformational domain to each amino-acid of a protein. 
This assignment leads to the detection of putative secondary structure elements and may be used to find specific conformational patterns in the entire PDB. 
The latter will be presented here through the example of EF-hands. These domains are composed of two helices separated by a 9-amino-acids loop known to bind calcium. 
They are important for signal transduction and muscle contraction \cite{LewitBentley:2000p234}.

%
\section*{Implementation}

RamaDA is programmed in the python (www.python.org) programming language, and employs the Biopython library \cite{Cock:2009p227}.
The online version of RamaDA is hosted on an Apache server.
An equivalent standalone version is also freely downloadable. 
Both take a protein structure file or a conformational pattern (see below) and and provide a graphic output of the analysis.

\subsection*{Statistical model}
Lovell et al. \cite{Lovell:2003p32} proposed a set of 500 protein structures extracted from the PDB to be representative of the statistical distribution of the $(\varphi, \psi)$ angles in the Ramachandran plot.
To this set, we added updated structures (PDB:1XFF, 1GOK, 1E70 and 1IG5), but kept one obsolete structure (PDB:1A1Y) in the list. This reference dataset contains 110\,018 amino-acids and is referred to throughout this manuscript as top500. This reference set was split into four subsets : glycines (Gly), amino-acids preceding a proline (pre-Pro), prolines (Pro) and the others (dataset called General). 

The seven conformational domains composing the Ramachandran plot that have been previously described in the literature (namely R-helices, L-helices, $\beta$, $\gamma$, $\zeta$, PPII and PPII$_R$) were fitted by a set of 2D-Gaussian functions cyclically defined over the complete periodic $[-180^{\circ}, 180^{\circ}]\times[-180^{\circ}, 180^{\circ}]$ domain.
Five parameters are necessary to describe each 2D-Gaussian : the position of the centre $(\varphi_{centre}$, $\psi_{centre})$, the standard deviations along both axes of the 2D-Gaussian $(\sigma_{\varphi^\prime}, \sigma_{\psi^\prime})$, and the angle made by the $\psi$ axis of the Ramachandran plot and the major axis of the 2D-Gaussian..
These parameters were first determined manually for each domain and then fitted to the top500 distribution assuming a Poisson noise.

\begin{figure}[!tpb]
  \begin{center} \includegraphics[width=0.45\textwidth]{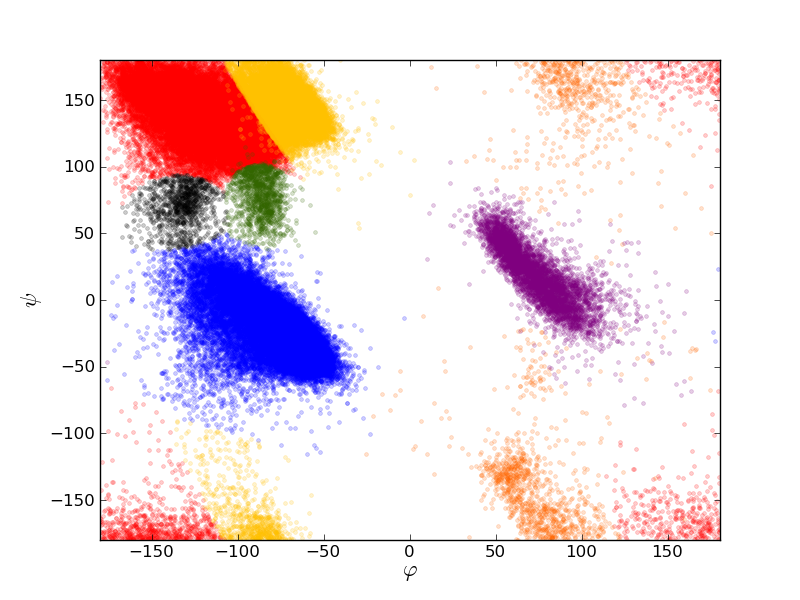}
     \caption{Ramachandran plot for top500.
     \label{fg-rama}
      Each dot represents an amino-acid. Seven domains are presented: $\beta$ in red, PPII in yellow, R-helices in blue, L-helices in violet, $\gamma$ in green, $\zeta$ in black and PPII$_R$ in orange.}
 \end{center}
\end{figure}

The statistical model of the Ramachandran plot implemented in RamaDA is composed of a set of 2D-Gaussian scaled to 1 (see Figure \ref{fg-rama}) 
and defined by the parameters found with top500. These parameters are gathered in 
Table \ref{tb-gauss}.  

\begin{table}[!tpb]
    \begin{tabular}{rcccc}
\hline
\textbf{Domain} & \textbf{($\varphi_c, \psi_c$)} & \textbf{$\sigma_{\varphi^\prime}$} & \textbf{$\sigma_{\psi^\prime}$} & \textbf{Angle}\\\hline
R-helices & (-63.07$^{\circ}$, -42.23$^{\circ}$) & 3.54$^{\circ}$ & 5.77$^{\circ}$ & -36.31$^{\circ}$\\
  & (-62.15$^{\circ}$, -28.74$^{\circ}$) & 12.00$^{\circ}$ & 4.69$^{\circ}$ & -61.76$^{\circ}$\\
  & (-83.72$^{\circ}$, -16.01$^{\circ}$) & 30.22$^{\circ}$ & 10.95$^{\circ}$ & -56.15$^{\circ}$\\\hline
L-helices (Gly)& (82.38$^{\circ}$, 6.89$^{\circ}$) & 7.55$^{\circ}$ & 20.63$^{\circ}$ & -32.81$^{\circ}$\\
         (pre-Pro) & (47.06$^{\circ}$, 5.93$^{\circ}$) & 5.58$^{\circ}$ & 6.79$^{\circ}$ & -27.38$^{\circ}$\\
        (General) & (56.35$^{\circ}$, 39.04$^{\circ}$) & 4.86$^{\circ}$ & 15.18$^{\circ}$ & -25.00$^{\circ}$\\\hline
$\beta$ & (-119.12$^{\circ}$, 136.48$^{\circ}$) & 15.77$^{\circ}$ & 29.98$^{\circ}$ & -52.51$^{\circ}$\\\hline
PPII & (-68.03$^{\circ}$, 144.89$^{\circ}$) & 9.65$^{\circ}$ & 16.67$^{\circ}$ & -30.37$^{\circ}$\\\hline
$\gamma$ & (-84.90$^{\circ}$, 69.28$^{\circ}$) & 5.85$^{\circ}$ & 10.82$^{\circ}$ & -6.51$^{\circ}$\\\hline
$\zeta$ & (-130.46$^{\circ}$, 76.31$^{\circ}$) & 5.90$^{\circ}$ & 12.80$^{\circ}$ & 12.25$^{\circ}$\\\hline
PPII$_R$ & (76.13$^{\circ}$, -162.12$^{\circ}$) & 11.75$^{\circ}$ & 41.02$^{\circ}$ & -29.27$^{\circ}$\\\hline
    \end{tabular}
    \caption {List of 2D-Gaussian functions needed for the statistical model of the Ramachanndran plot and their parameters}
    \label{tb-gauss}
\end{table}
\subsection*{Assignment and z-score calculation}

This set of 2D-Gaussian functions is used as a description of the backbone angle statistical distribution over the $[-180^{\circ}, 180^{\circ}]\times[-180^{\circ}, 180^{\circ}]$ space.
A particular conformation $(\varphi, \psi)$ is given a the probability of belonging to each of the seven conformational domains.
The most probable conformational domain is then assigned to this amino-acid at this location.

In the case of an NMR structure, which presented as a set of models, the same amino-acid can be assigned to various conformational states.
It is considered to be \emph{random-coil} if it is not found in the same domain for more than 65\% of the structures in the ensemble.
Similarly, a residue is considered to be \emph{extended} if the $\beta$ and PPII conformations are found in more than 65\% of the structures.

\begin{figure*}[!tpb]
  \begin{center} \includegraphics[width=0.77\textwidth]{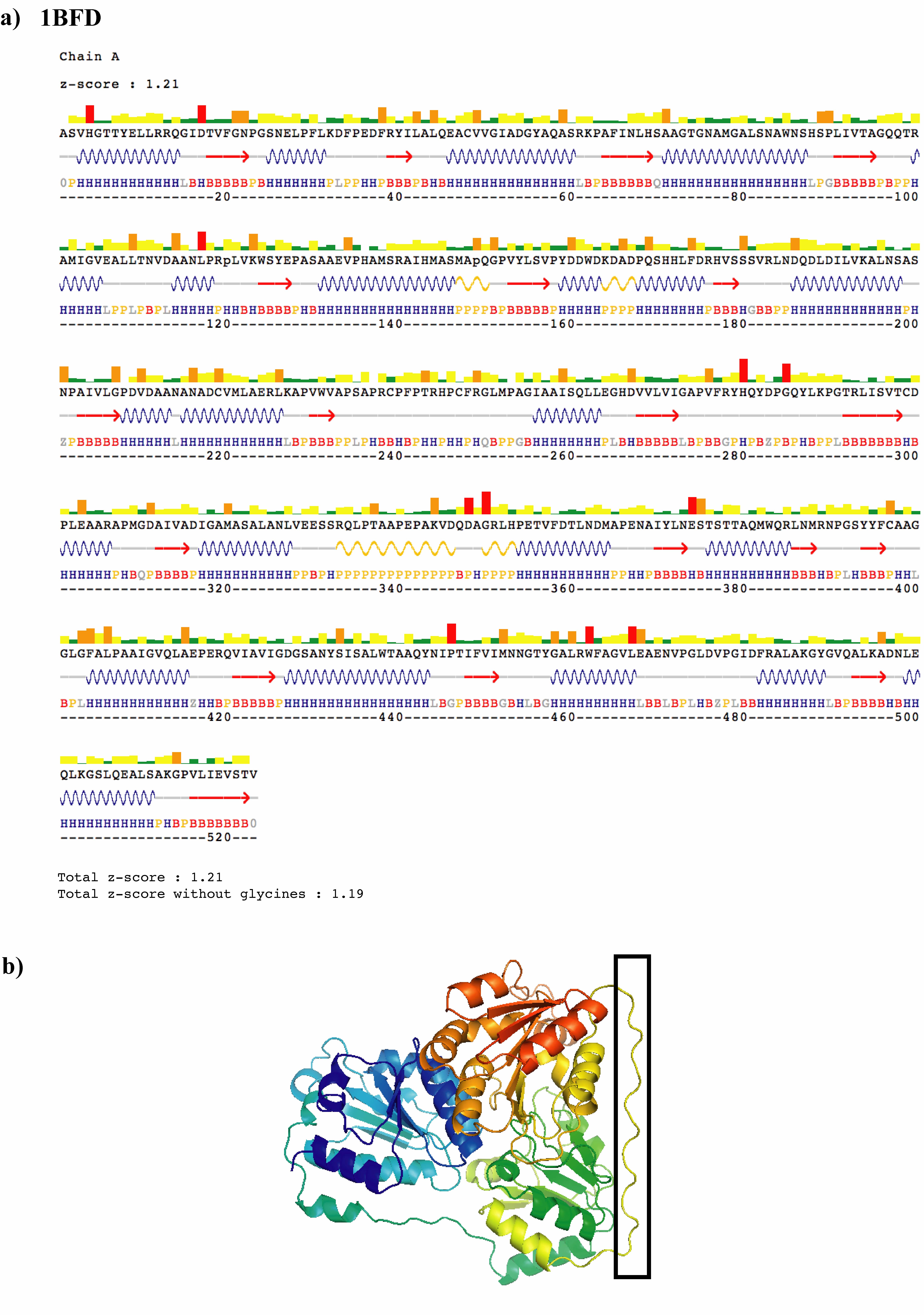}
    \caption{
      a) Output of RamaDA for PDB:1BFD.
      The first line shows the histogram of z-scores. Green is used for z-scores lower than 1, yellow for z-scores between 1 and 2, orange for z-scores between 2 and 3 and red for z-scores higher than 3.
      The second line corresponds to the protein sequence. Cis-prolines are detected thanks to their $\omega$ dihedral angle and indicated by \texttt{p}.
      The third line indicates putative secondary structure elements (blue waves for helices, yellow waves for PPII helices and red arrow for $\beta$ strands).
      The last line gives the conformational domains assignment (\texttt{H} for R-helices, \texttt{B} for $\beta$, \texttt{P} for PPII, \texttt{L} for L-helices, \texttt{G} for $\gamma$, \texttt{Z} for $\zeta$ and \texttt{Q} for PPII$_R$).
      b) 3-dimensional structure of the same protein. In the black box is highlighted the longest PPII helix found by RamaDA for this protein.
      This protein segment is indeed a PPII helix.}
      \label{fg-1BFD}
  \end{center}
\end{figure*}

For helices (respectively, $\beta$-strands and PPII helices), stretches of more than three residues assigned to the R-helices (respectively, $\beta$ and PPII) domain are indicated as putative secondary structure elements. Additionally, when a residue in the PPII conformation (respectively $\beta$) is found surrounded by 4 other amino-acids assigned to the $\beta$ conformation (respectively PPII) the secondary structure assignment is extended over this residue.
In order to highlight the potential secondary structure elements, RamaDA graphically displays indications of helices, $\beta$ strands and PPII helices along with the associated sequence.
A typical output of the RamaDA program is shown in Figure \ref{fg-1BFD}a for the \emph{Pseudomonas putida} benzoylformate protein (PDB:1BFD).

For each amino-acid conformation, a z-score is calculated as the distance from the centre of the assigned conformational domain related to the 2D-Gaussian standard deviations. 
The probability $P$ of a conformation is linked to the z-score $Z$ by the following equation :
\begin{equation}
P = e^{-\frac{Z^2}{2}}
\end{equation}

Formally, $Z$ can be positive or negative,
but the distance to the centre of the gaussian function is the only relevant parameter, a positive value of $Z$ is always found.
The z-score of a complete structure is the mean z-score of all the residues. A second z-score is calculated for all the amino-acids except glycines.
With these definitions, the mean z-score of a statistical series following a 1D-normal law is 0.7979  and 1.253 for a 2D-normal law.
As a consequence, one expects the mean z-score computed for a protein structure to be of the order of 1.25.
A larger value would indicate too many departures from the ideal geometry, while a smaller value would be the sign of too stringent constraints.

\subsection*{Conformational pattern recognition}

The RamaDA program was applied to the entire PDB, and a database containing the conformational assignment of each entry was made. This database is provided with the RamaDA program.
An updated version is provided weekly and can be downloaded from the website.

To look for a conformational pattern in this database, the RamaDA program uses regular expressions and gives the position of each result in the concerned protein. A scan over the entire PDB takes only several seconds on a desktop machine.

Patterns are described as a character string, using the syntax of regular expressions.
Conformational domains are represented by a letter : \texttt{H} for R-helices, \texttt{B} for $\beta$, \texttt{P} for PPII, \texttt{L} for L-helices, \texttt{G} for $\gamma$, \texttt{Z} for $\zeta$ and \texttt{Q} for PPII$_R$.
Conformational states \emph{extended} and \emph{random-coil} are represented by \texttt{e} and \texttt{R} respectively.

\section*{Results and discussion}

\subsection*{Secondary structure elements determination}

The RamaDA program was applied to the top500 set of proteins, and compared to the assignment given by the DSSP program \cite{Kabsch:1983p229}.
Table \ref{tb-secondary} 
gathers the results of this comparison.
\begin{table}[!tpb]
    \begin{tabular}[t]{ccc}
\hline
\textbf{RamaDA} & \textbf{DSSP} & \textbf{presence} \\
\textbf{domain} & \textbf{analysis} & \textbf{percentage} \\\hline
 \texttt{H} & $\alpha$ helix  & 64.0 \%\\
                & $\pi$ helix          &  $<$0.1 \%\\
                & $3_{10}$  helix  & \ \ 7.8 \%\\
                & extended strand & \ \ 1.4 \%\\
                & no assignment   & \ \ 4.8 \%\\
                & others                &  22.0 \%\\\hline
 \texttt{B} & $\alpha$ helix  &  $<$0.1 \%\\
                & $\pi$ helix         &    -  \\
                & $3_{10}$ helix  & $<$0.1 \%\\
                & extended strand & 64.9 \%\\
                & no assignment  & 23.1 \%\\
                & others                & 12.0 \%\\\hline
 \texttt{P} & $\alpha$ helix   & $<$0.1 \%\\
                & $\pi$ helix         &  -  \\
                & $3_{10}$ helix  & \ \ 0.4 \%\\
                & extended strand & 15.6 \%\\
                & no assignment   & 61.3 \%\\
                & others               & 22.7 \%\\\hline
    \end{tabular}
    \begin{tabular}[t]{ccc}\hline
\textbf{DSSP} & \textbf{RamaDA} & \textbf{presence} \\       
\textbf{analysis} & \textbf{domain} & \textbf{percentage} \\\hline      
$\alpha$ helix     & \texttt{H}     & 99.7 \%\\
                             & \texttt{B} & $<$0.1 \%\\
                             & \texttt{P} & $<$0.1 \%\\
                             & others     & \ \  0.3 \%\\\hline
$\pi$ helix          & \texttt{H} &   96.8 \%\\
                             & \texttt{B} &   -  \\
                             & \texttt{P} &   -  \\
                             & others     & \ \ 3.2 \%\\\hline
$3_{10}$ helix   & \texttt{H} & 93.4 \%\\
                             & \texttt{B} & \ \ 0.1 \%\\
                             & \texttt{P} & \ \ 1.2 \%\\
                             & others     & \ \ 5.3 \%\\\hline
extended strand           & \texttt{H} & \ \ 3.1 \%\\
                             & \texttt{B} & 85.1 \%\\
                             & \texttt{P} & \ \ 9.5 \%\\
                             & others     & \ \ 2.3 \%\\\hline
no assignment          & \texttt{H}  & 24.2 \%\\
                             & \texttt{B} & 34.0 \%\\
                             & \texttt{P} & 42.0 \%\\
                             & others     & 11.6 \%\\\hline
    \end{tabular}
    \caption{Comparison between assignments for secondary structure elements on top500.}
    \label{tb-secondary}
\end{table}

It can be seen that RamaDA finds a vast majority of the secondary structure elements found by DSSP.
Helices and extended strands found by DSSP are assigned in nearly 95\% of the cases to their respective conformational domains by RamaDA.
This is not surprising because secondary structure elements and conformational domains are strongly linked.
On the other hand, only about 72\% of the residues (respectively  65\%) found by RamaDA to lie in the helical domain (respectively $\beta$-domain) are described by DSSP as belonging to a helical secondary structures (respectively extended).
This comes from the fact that DSSP relies not only on backbone angles, but also on hydrogen-bound patterns which are not analyzed by RamaDA.
Short structural elements such as turns or loops and irregular regions are nevertheless constituted of amino-acids lying in regular conformational domain and are detected as such.

\begin{figure}[!tpb]
  \begin{center} \includegraphics[width=0.45\textwidth]{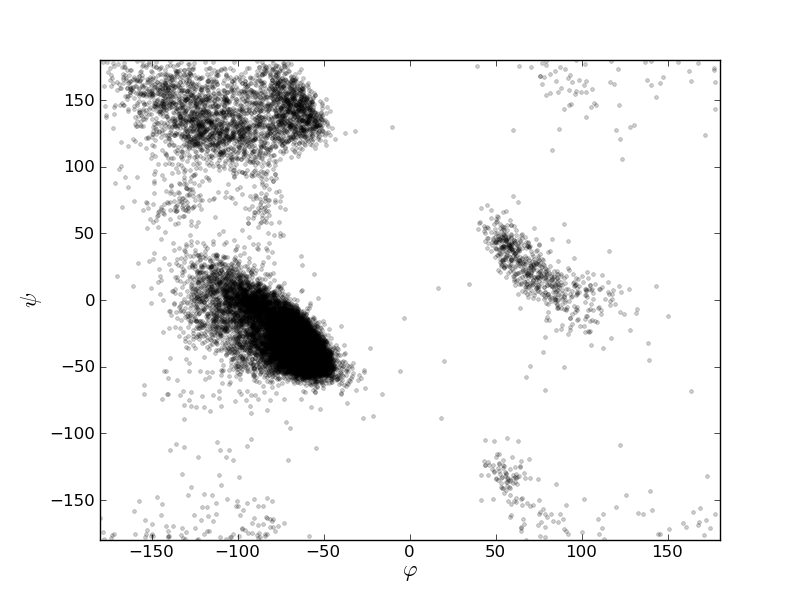}
    \caption{Ramachandran plot for top500 amino-acids assigned to helices according to DSSP}
    \label{fg-helix}
  \end{center}
\end{figure}

DSSP can also detect conformational irregularities in otherwise regular secondary structures.
Figure \ref{fg-helix} 
 illustrates this fact by presenting the Ramachandran plot of the helices found by DSSP on top500.
It can clearly be seen that while most of the amino-acids assigned by DSSP to a helical secondary structure lie in the helical conformational domain, many are scattered in all the accessible regions.
In contrast, all the residues assigned by RamaDA as being in a helical domain, lie inside the central helical domain of the Ramachandran plot.

Finally, DSSP does not attempt to analyze PolyProline II secondary structure (PPII), and most of these structures are classed as "no assignment" (Table \ref{tb-secondary}).
PPII is an extended regular secondary structure characterized by a pitch of three residues per turn and strong sequential side-chain contacts.
There is however no specific H-bound involved in the stabilization of the structural pattern.

In the PDB dated september 22$^{nd}$, 2011, 35715 chains of more than 5 residues and 590 chains of more than 10 residues were assigned a PPII conformational domain by the RamaDA program.
While a comprehensive analysis of the 590 segments has not been attempted, a rapid overview of some of the domains has confirmed the PPII assignment.
Such an example is shown in Figure \ref{fg-1BFD}b), where a 14 residues long canonical PPII helix was found.
This segment is not assigned to any secondary element in the sequence analysis given in the corresponding PDB entry, but is described as a PPII helix by the authors \cite{Hasson:1998fk}.

\subsection*{Example of EF-hands}
In this section, we use the conformational analysis provided by RamaDA to rapidly search the PDB for conformational patterns.
Many calcium-binding proteins contain a characteristic structure, called EF-hand, which is composed of a calcium-binding domain of 9 consecutive residues,
flanked by two $\alpha$-helices.
An ensemble of NMR structures for 4 EF-hands containing proteins was extracted from the PDB. 
These structures are those of calerythrin (PDB:1NYA), calmodulin (PDB:2BBM), parvalbumin (PDB:1RJV) and the human cardiac sodium channel NaV1.5 (PDB:2KBI).
As these proteins contains one to four EF-hand patterns, this set contains 10 EF-hand unique sequences.

The conformation assignments of the 10 calcium-binding loops were computed and compared and a consensus pattern was established.
Table \ref{tb-EFhand} 
compares three different consensus : amino-acids sequence, DSSP and RamaDA.

\begin{table*}[!tpb]
   \begin{tabular}{cccc}
\hline
                  & \textbf{Ca-binding} & \textbf{RamaDA}   & \textbf{DSSP}     \\
\textbf{proteins} & \textbf{domain}     & \textbf{analysis} & \textbf{analysis} \\
\hline
calerythrin (PDB:1NYA) & \texttt{DFDGNGALE} & \texttt{LHHLHLBBB} & \texttt{-SS--SSB-} \\
  & \texttt{DKNADGQIN} & \texttt{BHHLHLBBP} & \texttt{-SS--SEEE} \\
  & \texttt{DTNGNGELS} & \texttt{BHHLHLBBP} & \texttt{-TT-SSEEE} \\
  \hline
calmodulin (PDB:2BBM) & \texttt{DKDGDGTIT} & \texttt{BHHQHLBBB} & \texttt{-SSSS--B-} \\
  & \texttt{DADGNGTID} & \texttt{BHHLHLBBB} & \texttt{-SS-SSSB-} \\
  & \texttt{DKDGNGYIS} & \texttt{GHHQHLBBB} & \texttt{S-SSSSSB-} \\
  & \texttt{DIDGDGQVN} & \texttt{BHHQHLBBB} & \texttt{-SSS-SSB-} \\
  \hline
parvalbumin (PDB:1RJV) & \texttt{DKDKSGFIE} & \texttt{RHHLHLBBB} & \texttt{-TT-SS-B-} \\
 & \texttt{DKDGDGKIG} & \texttt{GHHLHLBPB} & \texttt{-SSSSSSB-} \\
 \hline
cardiac sodium channel (PDB 2KBI) & \texttt{DPEATQFIE} & \texttt{ZHHLHLBBP} & \texttt{-TT--SEEE} \\
\hline
consensus pattern  & \texttt{DX}[\texttt{DNE}][\texttt{GAK}]\texttt{X}[\texttt{GQ}]\texttt{X}[\texttt{ILV}]\texttt{X} & \texttt{.}\texttt{HHLHLBee} & [\texttt{-S}][\texttt{-ST}][\texttt{ST}][\texttt{-S}][\texttt{-S}][\texttt{-S}][\texttt{-SE}][\texttt{BE}][\texttt{-E}] \\\hline
    \end{tabular}
\caption{ Ensemble of 10 EF-hand structures from 4 different EF-hand containing proteins : sequence, RamaDA assignement and DSSP analysis. \texttt{X} et \texttt{.} correspond to any value, values between square brackets show the different possibilities for a same amino-acid and \texttt{e} is equivalent to [\texttt{BPe}].}
\label{tb-EFhand}
\end{table*}
It clearly appears that RamaDA offers a simple consensus conformational pattern.

The search for EF-hands domains was performed for the consensus pattern, flanked by two R-helices of at least 7 residues.
The consensus pattern used was \texttt{H\{7,\}.HH[LQ]HLBeeH\{7,\}}
(using the regular expression syntax, where a dot matches everything, \texttt{A\{x,\}} means \emph{A repeated at least x times}, \texttt{[AB]} means A \emph{or} B and using the nomenclature defined in the implementation section).
Using this pattern, 537 hits were found in the PDB dated september, 22$^{nd}$ 2011.

Each one of these hits was manually analyzed to confirm the presence of a EF-hand domain, using either direct description given by the authors or known calcium binding activity. 523 hits were true positive, meaning the chosen pattern allows the detection of EF-hand domains with a 97.3\% accuracy.

The chosen consensus pattern may appear too strict, especially concerning the flanking helices. As it was stressed previously, some amino-acids of helices may not adopt a R-helices conformational domain. Then, the helices' boundaries may be different than expected.

The pattern  \texttt{H\{6,\}..HH[LQ]HLBeeH\{7,\}} leads to the detection of three more PDB files that are also true positives. However, the pattern \texttt{H\{7,\}.HH[LQ]HLBee.H\{6,\}} leads to the detection of three true positive hits out of 15 additional files, letting the accuracy drop to 96.2\%.

To compare these results to others, a search for the amino-acid sequence consensus was performed with PATTINPROT \cite{Combet:2000p2534} over the entire PDB.
932 PDB files can be found and only 224 of them are also found by RamaDA meaning that more than 300 files found thanks to the conformational pattern are missed with the amino-acid sequence pattern. 
Moreover, a vast majority of the files detected by PATTINPROT only are false positives.
As for the DSSP consensus, no search was performed as it is obviously too large to give accurate results.

\section*{Conclusion}
The description of a protein given by RamaDA is a synthetic view of the local protein chain organization.
It provides an accurate detection of secondary structure elements and also local patterns adopted by amino-acids.
This information is rapidely obtained, and was made available on internet as well as a standalone application.
Moreover, the example of EF-hands shows that the use of RamaDA enhances conformational pattern recognition.
Through this example, a promising approach of structural biology is developed.
It is also interesting to note that, thanks to its simple usage and its fast results, RamaDA can be applied on large sets of structure files including, for example, those created via multiconformational analysis.

\section*{Acknowledgements}
We would like to thank the CNRS and NMRtec for their financial support, Roland Stote and Bruno Kieffer for fruitful discussions and Claude Ling for the technical support on the website.
This study was partially supported by funds from the Agence Nationale de la Recherche (ANR-07-PROTANIN)

\section*{List of abbreviations}
RamaDA: Ramachandran Domain Analysis ; PDB: Protein Data Bank

\bigskip

\bibliographystyle{hacm}  
\bibliography{RamaDA_arXiv}     


\end{document}